\documentclass[conference]{IEEEtran}
\usepackage{amsthm}
\usepackage{amsfonts}
\usepackage{mathtools}
\usepackage{mathrsfs}
\usepackage{xspace}
\usepackage{amsmath}
\usepackage{tabularx}
\usepackage{cite}

\begin{document}
\title{A Physical Testbed for Intelligent \\ Transportation Systems}

\author{\IEEEauthorblockN{Adam Morrissett, Roja Eini, Mostafa Zaman, Nasibeh Zohrabi, Sherif Abdelwahed}
\IEEEauthorblockA{Department of Electrical and Computer Engineering \\
Virginia Commonwealth University \\
Richmond, Virginia 23220 \\
Emails: morrissettal2@vcu.edu, einir@vcu.edu, zamanm@vcu.edu, zohrabin@vcu.edu, sabdelwahed@vcu.edu}}

\maketitle

\begin{abstract}
Intelligent transportation systems (ITSs) and other smart-city technologies are increasingly advancing in capability and complexity.
While simulation environments continue to improve, their fidelity and ease of use can quickly degrade as newer systems become increasingly complex.
To remedy this, we propose a hardware- and software-based traffic management system testbed as part of a larger smart-city testbed.
It comprises a network of connected vehicles, a network of intersection controllers, a variety of control services, and data analytics services.
The main goal of our testbed is to provide researchers and students with the means to develop novel traffic and vehicle control algorithms with higher fidelity than what can be achieved with simulation alone.
Specifically, we are using the testbed to develop an integrated management system that combines model-based control and data analytics to improve the system performance over time.
In this paper, we give a detailed description of each component within the testbed and discuss its current developmental state.
Additionally, we present  initial results and propose future work.
\end{abstract}


\IEEEpeerreviewmaketitle

\section{Introduction}
\label{sec:introduction}
Successful implementations of intelligent transportation systems (ITSs) and other smart city systems require numerous subsystems to be integrated together.
At the same time, the system as a whole needs to satisfy safety and performance specifications.
Many systems are designed and analyzed using simulation tools, but, to the best of our knowledge, no single software suite is capable of adequately simulating all aspects of such large-scale systems.
Instead, researchers typically combine separate modeling, network simulation, and traffic simulation software.
Not only is the integration tedious and complicated, but it is also a relatively rigid work-flow; therefore, we propose a software- and hardware-based testbed for ITS research.

The current availability of ITS testbeds is limited, with many focusing on enabling wireless sensor network (WSN) research [1], [2].
While these testbeds provide guidance on designing and deploying a complex testbed, they do not incorporate control of ITSs or smart buildings.
Additionally, the WSN testbeds often deploy to real-world cities, which may be infeasible for certain researchers due to economic or regulatory constraints.
The closest related work to ours involves researching autonomous driving with fleets of small-scale toy cars [3].
Their results successfully demonstrate the viability of using a low-cost, small-scale testbed for both egocentric and cooperative ITS research.
This testbed, however, is focused only on autonomous vehicles, whereas our system includes it as part of a wider range of capabilities including, for instance, a traffic control and management system [4].
The authors in [5] propose a cloud-based microservices architecture for ITS testbeds, but it too is limited to connected vehicles.

Our testbed consists of a network of connected vehicles, a network of intersection controllers, data analytics services, and a dashboard.
Vehicles can be controlled either manually by a human operator or autonomously by a variety of model-based or machine learning\textendash based control systems.
Additionally, each vehicle shares its location information with a database service using an infrastructure-based wireless communication technology.

A distributed traffic-management system is then able to reconfigure the parameters within the intersection controllers to optimize the traffic flow in the network.
Centralized model-based approaches have been widely employed to manage traffic networks, subject to various disturbances such as the vehicle flow, turning rate, timing, incidents, climatic conditions, and road structure conditions [6]-[11].
Since traffic networks are innately networked systems comprised of several lanes with different structures and are associated with large-scale optimization problems, using a centralized strategy demands an extensive computation overhead as well as complex communications. 
Authors in [12] and [13] address the large-plant optimization problem using the distributed control schemes. In [12], a distributed model-based controller is designed for the heating/cooling system of a large building considering the environment factors, and in [13], a distributed control structure is proposed for a medium-voltage DC shipboard power system for lower computational overhead and higher flexibility. 
In this work, we design a distributed control system that utilizes the traffic flow and disturbances predictions to optimize the control inputs (signal split timings) in each intersection.
Additionally, traffic patterns and system environments can change as time progresses.
We therefore include a data analytics service that uses machine learning to aggregate all of the data in the testbed and generate model updates for all the model-based controllers.

Finally, researchers and city administrators need to be able to view large and complex data in an intuitive representation.
The dashboard aggregates testbed data and generates graphs or other visual elements.
Additionally, it provides an interface through which system parameters can be manually changed.
This allows researchers to easily modify the testbed from a centralized location.
It also enables city administrators to manually override certain systems, like the traffic management system, to accommodate special events.

The rest of the paper is organized as follows.
Section \ref{sec:system_structure} describes the testbed and all of the subsystems it comprises. 
Section \ref{sec:project_objectives} discusses its different objectives and application areas. 
Section \ref{sec:implementation} presents the initial implementation and simulation results. 
Finally, Section \ref{sec:conclusion} provides the conclusions and discusses future research.

\section{System Structure}
\label{sec:system_structure}
\begin{figure}
    \centering
    \includegraphics[width=\columnwidth]{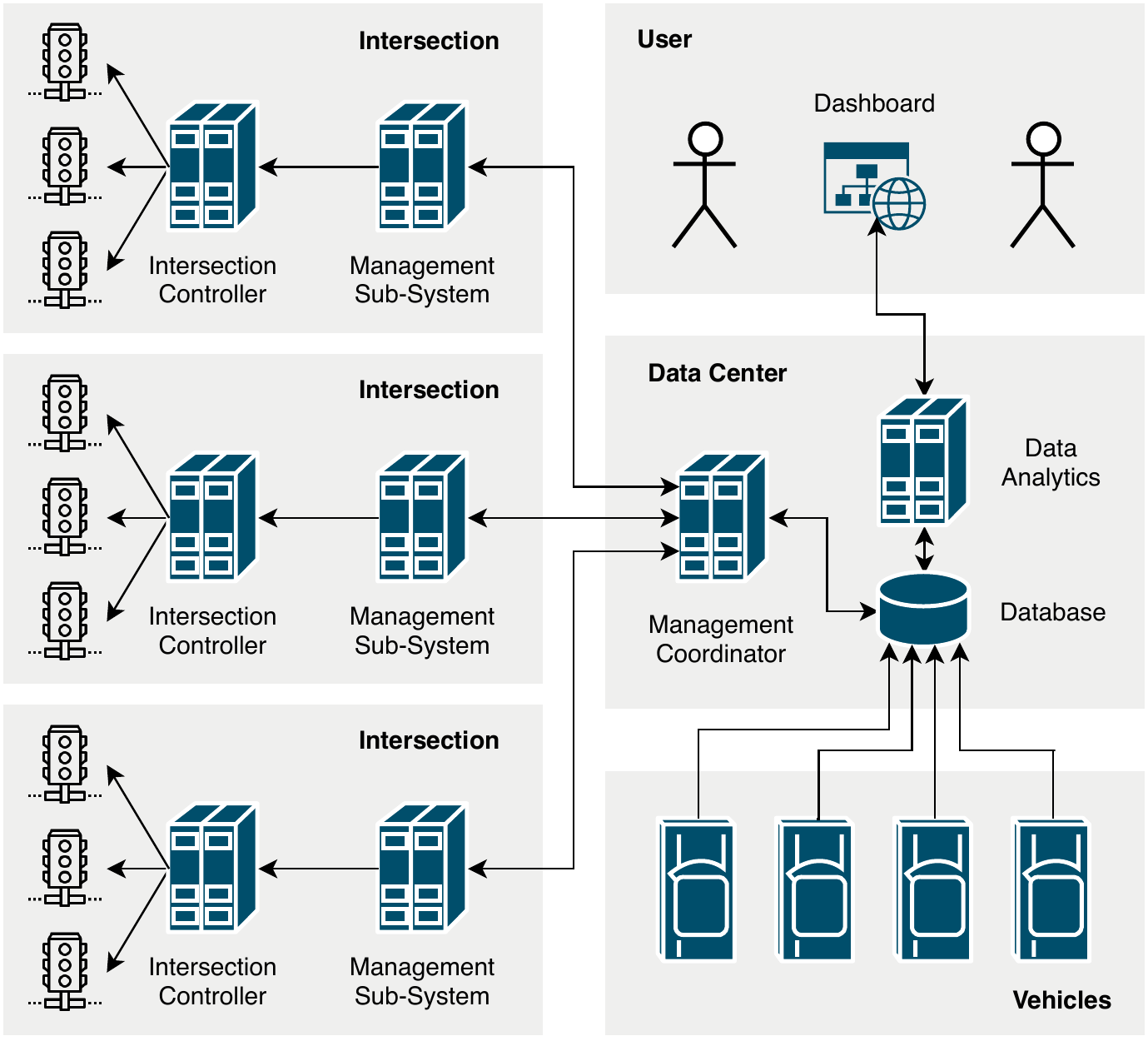}
    \caption{System-wide diagram}
    \label{fig:system_diagram}
\end{figure}

In this section, each of the systems that composes the testbed is described.
Fig. \ref{fig:system_diagram} shows a diagram of the entire testbed and how each of the systems connects.
Vehicles in the \textit{Vehicles} area and intersection controllers in each of the \textit{Intersection} areas send their location and traffic-flow data, respectively, to the database located in the \textit{Data Center} area.
The traffic management coordinator, also in the \textit{Data Center} area, uses this information to globally optimize the flow of traffic.
Users are able to interact with the testbed through the dashboard, which is shown in the \textit{User} area.
The dashboard provides data visualization and manual overrides for system parameters throughout the testbed.

\subsection{Vehicle System}
The \textit{Vehicles} area, shown in Fig. \ref{fig:system_diagram}, comprises all of the vehicles on the road network, including both operated and parked vehicles.
Each vehicle is equipped with both sensor and communication suites, and they are either manually or autonomously controlled.
When vehicles are in the manual mode, the sensors and a controller maintain the desired speed and turn angle inputted by a human driver.
In the autonomous mode, however, the vehicles are controlled by a machine learning algorithm that inputs video from a front-facing camera and outputs the appropriate speed and turn-angle values.
During their operation, all of the vehicles transmit their state information, such as current speed, turn angle, location, etc., to the database.

\subsection{Intersection Controllers}
An intersection controller, shown in the \textit{Intersection} areas of Fig. \ref{fig:system_diagram}, is responsible for safely regulating the traffic phases at a local intersection.
At its core, the controller is modeled as a state machine with each state corresponding to a particular sequence of phases; the state-transition time is a parameter.
Each intersection controller transmits its current state and parameters to the database.
These parameters can be modified by an external traffic management system or by users through the dashboard.

Additionally, the intersection controllers can be equipped with a variety of external peripherals, such as a sensor suite to detect and monitor vehicles or a communication suite to communicate directly with vehicles.

\subsection{Traffic Management System}
The traffic management system is comprised of a network of vehicle systems, a network of intersection controllers, and a distributed traffic controller; all of which are shown in Fig. \ref{fig:system_diagram}.
The distributed traffic controller consists of local controllers that integrate with the intersection controllers, shown in the \textit{Intersection} areas, and a centralized coordinator, shown in the \textit{Data Center} area.
Vehicles transmit their location information to the database, and this location information is collected by the coordinator.
After it globally optimizes the vehicle routes, the local controllers update the timer parameters of their respective intersection controllers to optimize the traffic congestion in the roads.

\subsection{Database}
The database, shown in the \textit{Data Center} area of Fig. \ref{fig:system_diagram}, serves as the hub for all system state information in the testbed.
Because the network and service architectures are designed to be modular, future data aggregation, manipulation, and visualization services can be easily connected to the database.
The modularity also allows for the network to be easily secured by implementing firewalls, demilitarized zones, and other security measures.
Furthermore, the database can be distributed in the future to increase data resiliency and fault tolerance.

\subsection{Data Analytics}
The data analytics service, shown in the \textit{Data Center} area of Fig. \ref{fig:system_diagram}, is responsible for aggregating relevant data to improve existing models for controllers.
A goal for the testbed is to use it during the development of novel control systems that utilize both model-based control and machine learning.
As part of this, the data analytics service reads all of the data collected on the environment for a specific system and generates an update to the existing environmental model.
This will allow control systems to better learn about and adapt to their environment.

\subsection{Dashboard}
The dashboard, shown in the \textit{Users} area of Fig. \ref{fig:system_diagram}, presents users with a visualization of data from the database.
For a traffic management system, visualization can include traffic density over periods of time.
Data visualization provides an intuitive representation of information with which city administrators can make informed decisions regarding city operations.

Additionally, users can view and change system parameters from the dashboard.
This includes manually setting phase timing on the intersection controllers or restricting access to certain roads.
As a testbed, being able to change parameters from a single interface allows researchers to easily control multiple systems without having to individually connect to each of them.
As a deployed system, it would allow city administrators to modify the normal operating parameters for special events that are not part of everyday operations.

\subsection{Network Infrastructure}
All of the systems and services communicate via a traditional, infrastructure-based, network, and the network connections are shown as lines in Fig. \ref{fig:system_diagram}.
The services communicate to each other through a high-bandwidth wired backbone network.
The vehicles, intersection controllers, and similar systems communicate to the services through a wireless access point.

\section{Project Objectives}
\label{sec:project_objectives}
This section describes the main objectives of the proposed traffic testbed.
They can be broadly categorized into research and education.

\subsection{Integrated Control and Data Analytics Management}
The main objective of the testbed is to use it for new integrated control systems that utilize both model-based controls and machine learning.
With traditional model-based controls, controllers operate on systems based on the models of system and its environment.
These models, however, can be subject to change over time, which requires deriving a new model.
By using machine learning, a controller can be developed with an initial model, and the model can be continuously improved.

\subsection{Traffic management Systems}
The control objective in traffic management is to minimize both the total time spent (TTS) traveling by the vehicles, and the traffic congestion in the roads.
Additionally, the traffic management system should be robust against model disturbances due to a variety of reasons including sharp inflow rate changes, incidents, and environmental disturbances.

With the previously-described integrated approach, a generic model of traffic flow can be used to develop the traffic management system effective for all environments.
When this system is deployed, machine learning would monitor the environment to optimize the system for its specific environment.
Additionally, the system would be able to adapt to trends in the traffic flow.

Other projects include developing a distributed traffic management system that does not seize when a maintenance or failure happens in one part of the system or its controller. 
The traffic system and its control scheme are developed to be  modular and generic; e.g., the system-control goals or the traffic structure can be customized based on the city administrators' preferences. 
A modular prototype of the traffic system also can be used for further experimental purposes in the area of ITSs’ features and challenges (e.g., studying the effects of actuators' failures or system maintenance in a traffic network).

\subsection{Intelligent Vehicle Systems}
The testbed will provide a rich development environment for a variety of intelligent vehicle systems.
Vision-based environment perception is a complex task that is difficult to develop using only simulations.
By using physical, scale-model, vehicles, and roads, researchers can quickly create complex environments with relative ease.

Additionally, multiple systems, communication infrastructures, and computing services can be easily combined without having to tediously integrate multiple simulation software suites.
This will prove especially useful for connected autonomous vehicles and vehicle platooning.

\subsection{Education}
Large-scale intelligent systems are innately multi-disciplinary, and as such, they provide the opportunity for large collaborations.
For example, safely and reliably controlling the systems requires expertise in controls engineering; creating a communication infrastructure requires expertise in network engineering; and data aggregation and visualization require expertise in machine learning.

For many universities and students, beginning research on ITSs and smart-cities can be challenging.
By constructing a physical testbed, students can interact with individual systems to further their understanding of control systems.
This testbed provides opportunities for students to learn new technologies in the promotion of future smart cities.

\section{Implementation}
\label{sec:implementation}
This section describes the current implementation of the traffic testbed.

\subsection{Vehicle Control}
\begin{figure}
    \centering
    \includegraphics[width=.75\columnwidth]{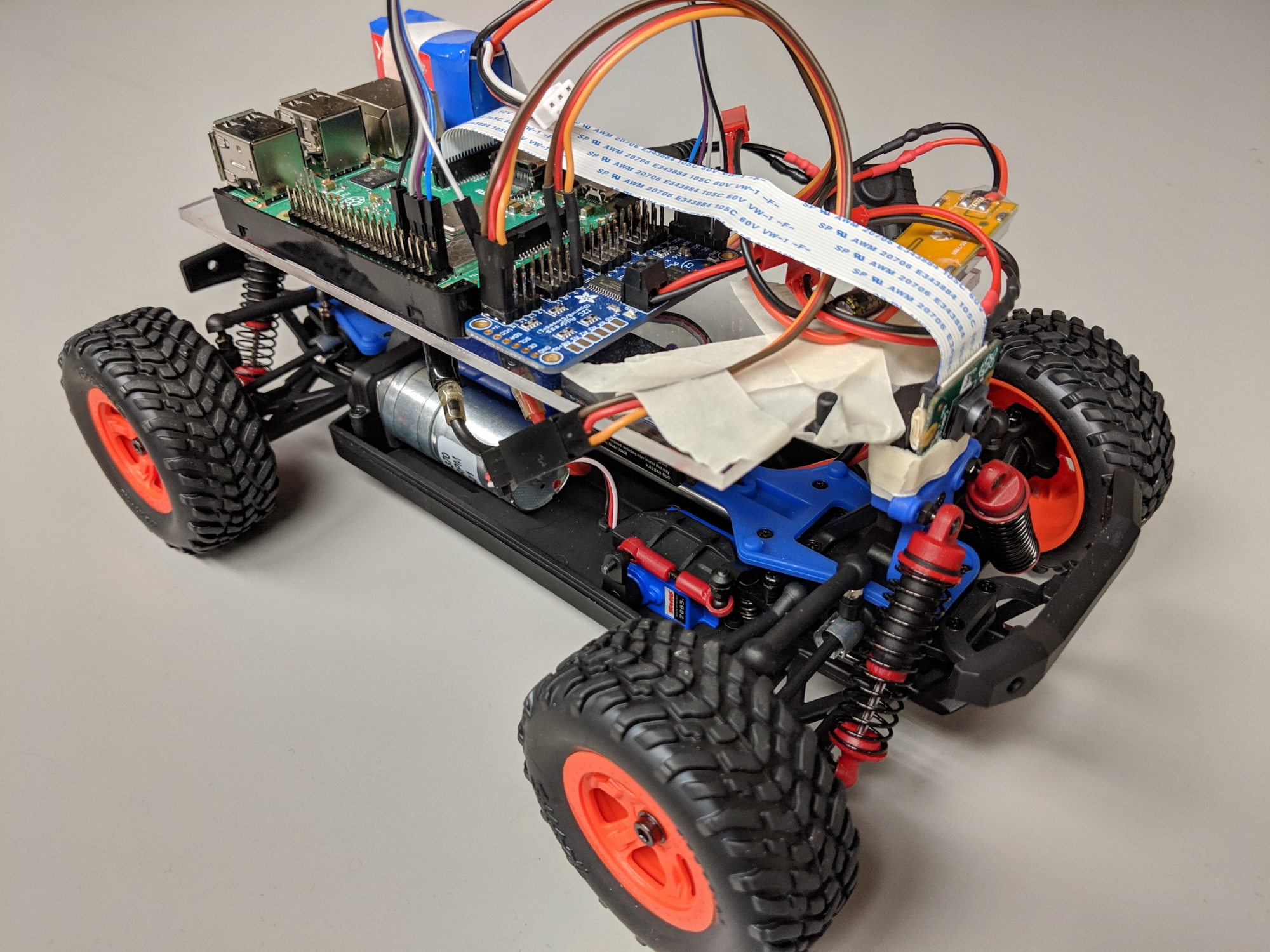}
    \caption{Autonomous vehicle implementation}
    \label{fig:autonomous_vehicle}
\end{figure}

The vehicle platform, shown in Fig. \ref{fig:autonomous_vehicle}, consists of a LaTrax Desert Prerunner 1/18-scale radio-controlled (RC) car chassis, an electronic speed controller (ESC), a Raspberry Pi 3 Model B+, and a Raspberry Pi Camera Module v2.
It is an adaptation of the DeepPicar [14] but with a focus on integration into a vehicular network rather than evaluating the viability of embedded systems in autonomous vehicles.
In addition, our platform uses a smaller-scale chassis and currently off-boards autonomous control to a remote server.

\begin{figure}[ht]
    \centering
    \includegraphics[width=\columnwidth]{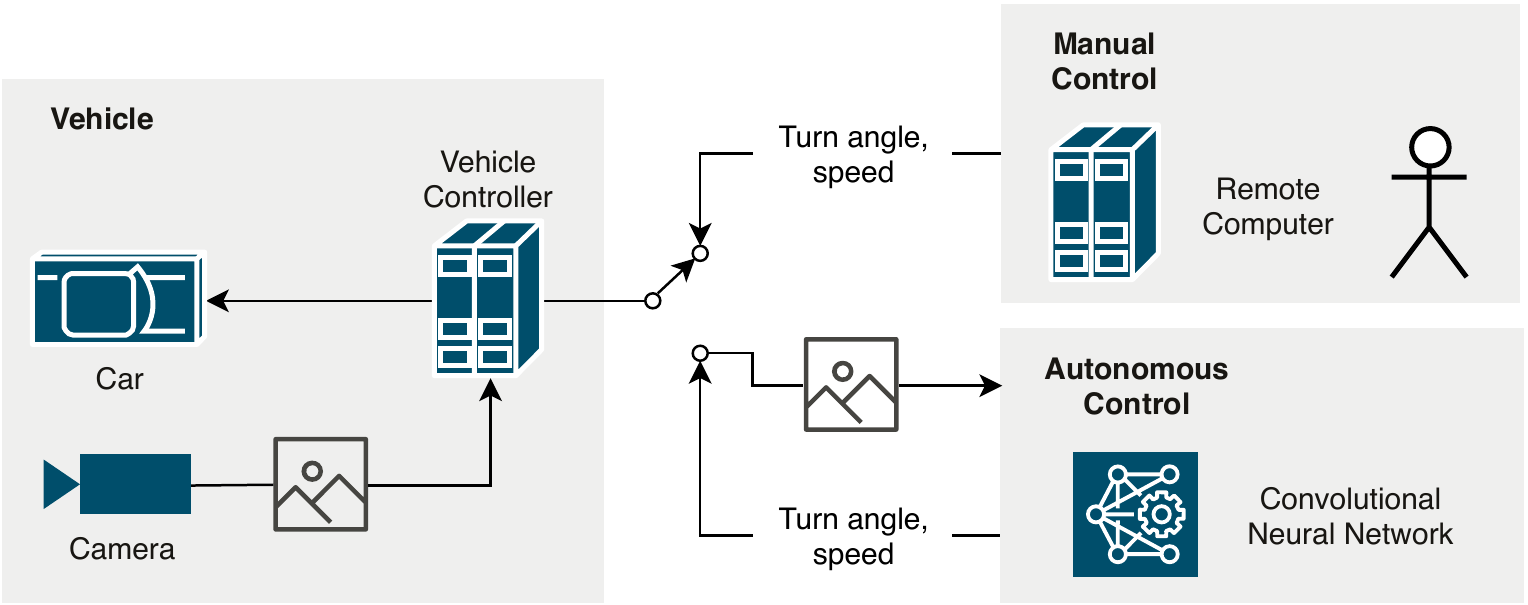}
    \caption{Vehicle control diagram with two modes: manual and autonomous}
    \label{fig:autonomous_vehicle_diagram}
\end{figure}

The vehicle has two control modes, which are shown in Fig. \ref{fig:autonomous_vehicle_diagram}: manual and autonomous.
When in the manual control mode, the vehicle receives turn-angle and speed values from a remote computer.
This computer interfaces with an FrSky Taranis Q transmitter and streams the commands to the vehicle.

In the autonomous control mode, the control flow consists of three main steps: front-view image streaming to a remote convolutional neural network (CNN) service, CNN-based image processing, and generated turn-angle and speed value streaming to the vehicle.
This flow is repeated for the operational lifetime of the vehicle.
The CNN architecture is based on the one used in [15], and its parameters are given in Table \ref{tab:cnn}.
It is trained using end-to-end learning [15],[16].
\begin{table}[b]
    \centering
    \caption{CNN Network Parameters}
\begin{tabular}{cccc}
	\hline
	Layer & Feature Map & Kernel Size & Stride \\
	\hline
	Input & $200 \times 66 \times 3$ & - & - \\
	Normalization & $200 \times 66 \times 3$ & - & -  \\
	Convolutional 1 & $24$@$98 \times 31$ & $5 \times 5$ & $2 \times 2$ \\
	Convolutional 2 & $36$@$47 \times 14$ & $5 \times 5$ & $2 \times 2$ \\
	Convolutional 3 & $48$@$22 \times 5$ & $5 \times 5$ & $2 \times 2$ \\
	Convolutional 4 & $64$@$20 \times 3$ & $3 \times 3$ & None \\
	Convolutional  5 & $64$@$18 \times 1$ & $3 \times 3$ & None \\
	Flatten & $1\,164$ & - & - \\
	Fully Connected 1 & $100$ & - & - \\
	Fully Connected 2 & $50$ & - & - \\
	Fully Connected 3 & $10$ & - & - \\
	Output & $1$ & - & - \\
	\hline
\end{tabular}
    \label{tab:cnn}
\end{table}

The vehicle communicates to the control servers and database through the IEEE 802.11n/ac standard and User Datagram Protocol (UDP) packets.
We are using the current configuration to test the scalability of the system and see how it performs as more cars are added to the network.
This will let us determine the feasibility of off-boarding certain decision-making processes in near-real-world scenarios.

Our future plans involve replacing the Raspberry Pi with the newly released NVIDIA Jetson Nano Developer Kit and on-boarding the CNN controls.
Having a graphics processing unit (GPU) in an embedded platform that also has the same interfacing options as the Raspberry Pi will allow us to implement more complex algorithms and optimize them without sacrificing input-output (I/O) capabilities.

\subsection{Intersection Controller}
\begin{figure}[ht]
    \centering
    \includegraphics[width=.75\columnwidth]{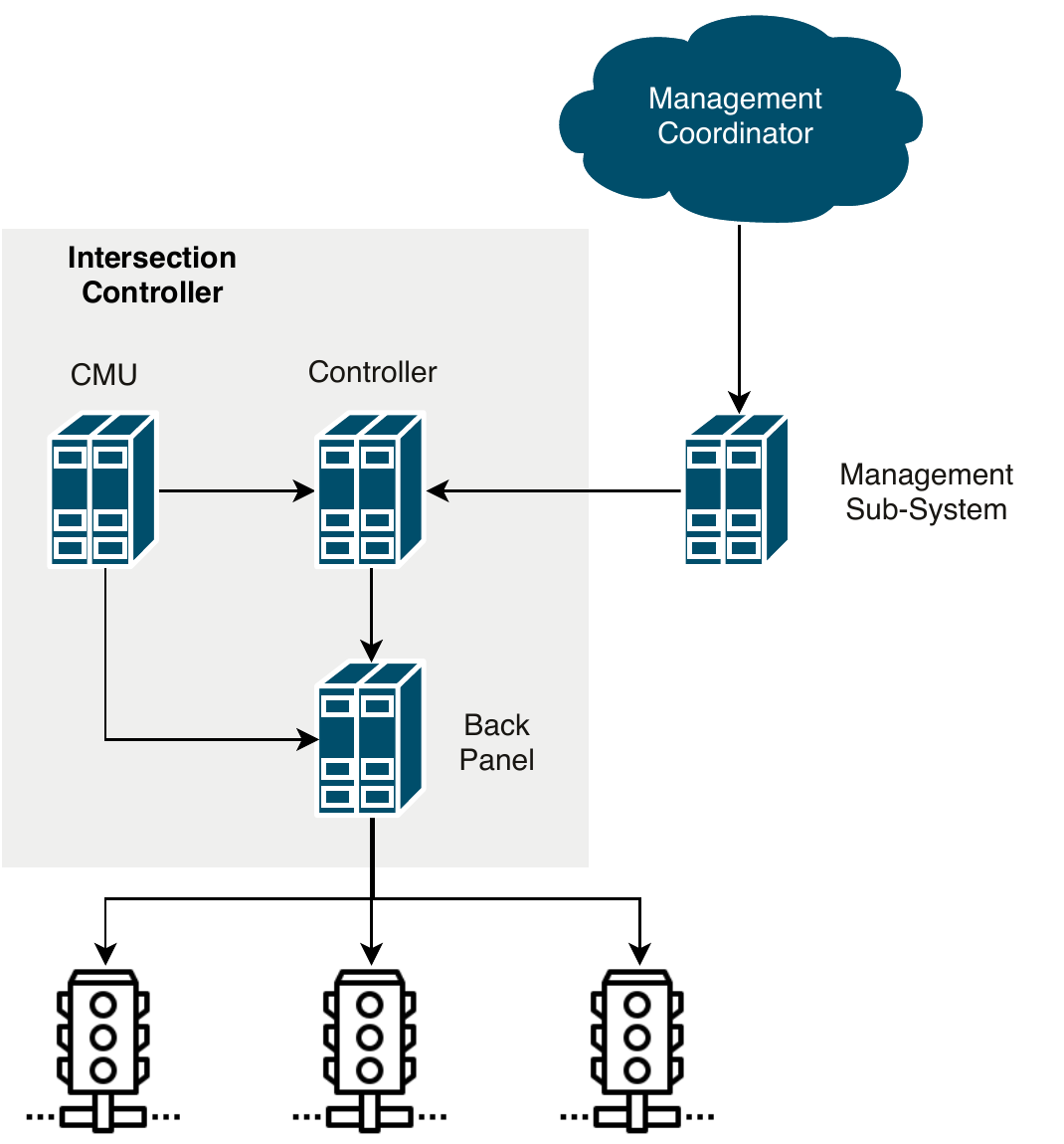}
    \caption{Intersection controller implementation architecture}
    \label{fig:intersection_controller_diagram}
\end{figure}

The architecture of the intersection controllers, shown in Fig. \ref{fig:intersection_controller_diagram}, is based on the NEMA TS 2 Type 2 [17] standard.
It comprises the following units: a main logic controller, a conflict management unit (CMU), and a back panel.
Additionally, we have added an interfacing module that connects it to the networking infrastructure.

The main logic controller is responsible for controlling which phases are enabled at any given time.
Under normal circumstances, it cycles through all of its programmed phase sequences based on a timer, and this timer parameter can be modified through the dashboard or through the traffic management system.
The logic controller also interfaces with the CMU, which allows for non-standard phase sequences and improved reliability.

When privileged vehicles, such as emergency vehicles, approach the intersection, the CMU overrides the current phase sequence in the controller to allow the privileged vehicle to enter the intersection.
The CMU also monitors the output of the main logic controller and will override its output if incompatible phases are enabled at the same time.
In this event, the CMU will force the intersection lights to a pre-configured state, such as all red lights flashing.

The back panel serves as the light interface for the controller and the CMU.
When the controller changes phases, it sends the updated list of phase states to the back panel, which then switches the corresponding lights on or off.
If conflicting phases are detected or a privileged vehicle is approaching, the CMU overrides the output of the controller through the back panel.

The interfacing module provides network connectivity to the city-wide traffic management system.
This unit transmits the controller's current state information to the database, and it receives parameter change commands from the traffic-management system.
When parameter update commands are received, it forwards the new values to the controller unit.

\begin{figure}[ht]
    \centering
    \includegraphics[width=.7\columnwidth]{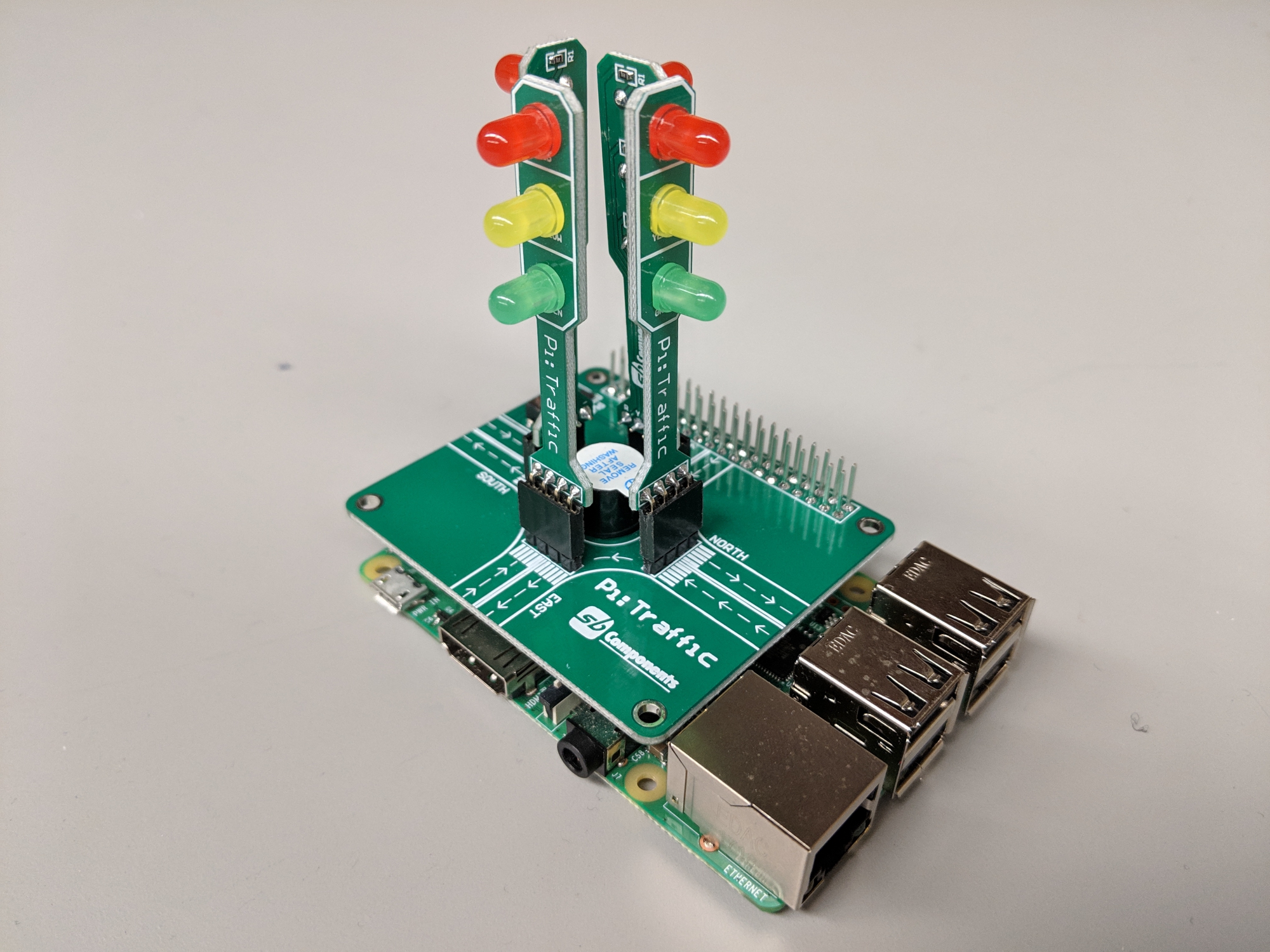}
    \caption{Local intersection controller implementation on the Raspberry Pi 3 Model B$+$}
    \label{fig:rpi_controller}
\end{figure}

The intersection controller is implemented on a Raspberry Pi 3 Model B$+$ using the Python programming language, and the physical platform is shown in Fig. \ref{fig:rpi_controller}.

Each of the subsystems is contained within a separate thread, and they communicate with each other using Transmission Control Protocol (TCP) packets.
By using threads and TCP communication, each subsystem can be easily migrated to a physically-separate system if needed.

\subsection{Traffic Management System}
\begin{figure*}[htpb]
	\centering
	\includegraphics[width=0.9\textwidth]{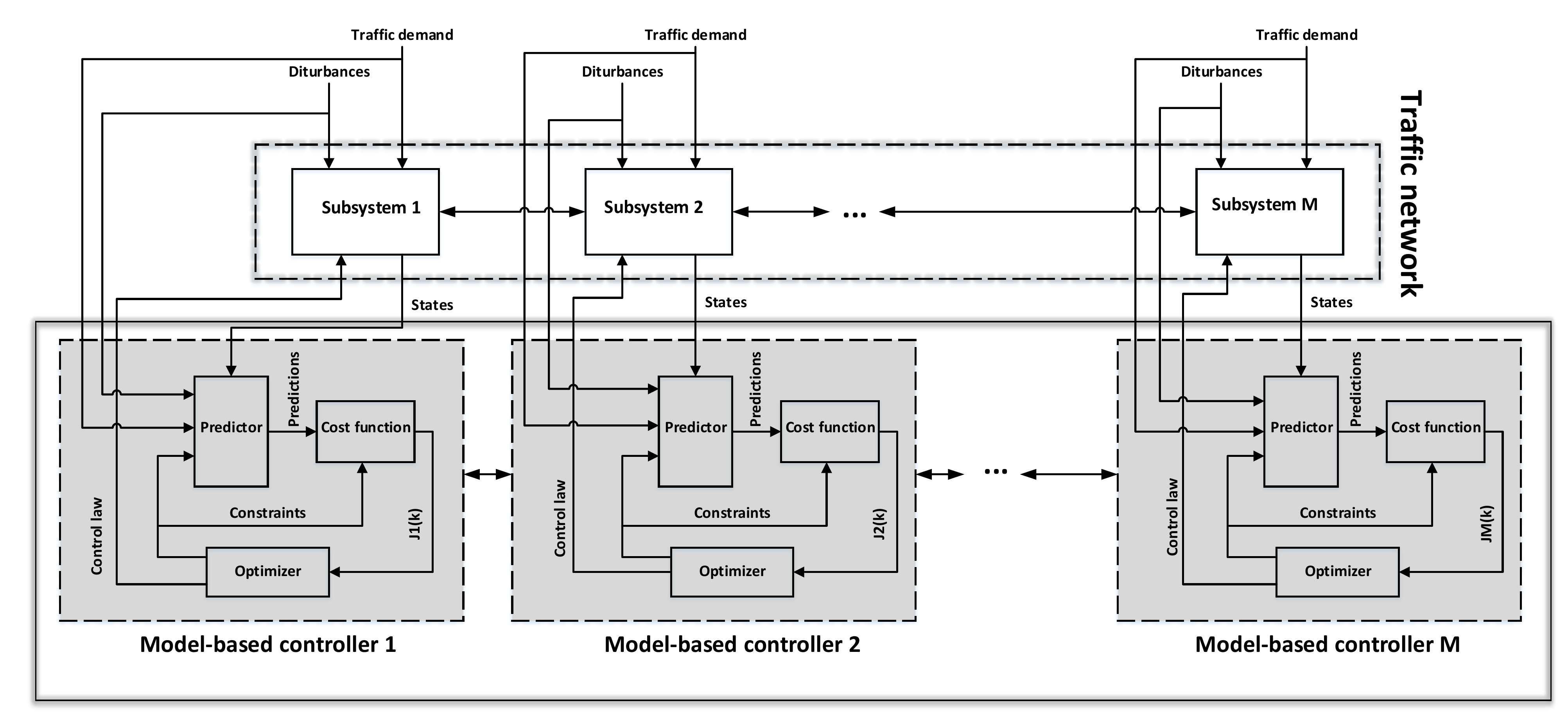}
	\caption{Distributed model-based control of $M$ interacted subsystems}
	\label{fig:MPC_dis}
\end{figure*}

The traffic management system is simulated in MATLAB. The final aim is to implement the simulated traffic management system in the testbed. 
The traffic controller is developed distributed; e.g., one local model-based controller is designed for each subsystem.
The subsystems are the traffic network intersections as shown in Fig. \ref{fig:system_diagram}. 
Based on the local data provided by the connected vehicles, intersection sensors, or city administration preferences, all the sub-controllers coordinate to optimize the traffic flow in all the roads to achieve the network's global objective.
Fig. \ref{fig:MPC_dis} shows a diagram of a distributed model-based control for a network with $M$ subsystems and intersections, where $M$ is the number of all subsystems.
 
For the distributed control structure, the traffic network is decomposed into $M$ subsystems. 
A center junction (also termed \textit{intersection}) with the connecting roads is considered as a subsystem; the subsystems interact through the data center (Fig. \ref{fig:system_diagram}). The interaction variables among subsystems are the inflows and outflows from the upstream and downstream intersections, respectively. 
The flow rates are considered as the Gaussian functions in the simulations. 
The problem operating constraints include the model constraints, the demand constraints, the input constraints, and the state constraints. 
At each time step of the simulation, the local constrained optimization problem is solved for each subsystem (center intersection) using the CasADi optimization tool [18] in MATLAB. 
Minimizing the total travel time spent (TTS) criterion, corresponding to the accumulated amount of time spent by the vehicles, and the number of vehicles are the control objectives.
The traffic controller generates the optimum control inputs at each step. 
The inputs are then transmitted to the intersection's controller to control the signal split time.
Fig. \ref{fig:intersection_controller_diagram} shows how the traffic-management system interfaces with the intersection controller.

At their core, intersection controllers are state machines that regulate traffic phases based on a timer.
More advanced intersection controllers can override the default traffic pattern by detecting vehicles in specific lanes, such as turn lanes, or privileged vehicles, such as emergency vehicles.
Others can interface with sensor suites or remote systems to monitor traffic flow or receive additional control commands, respectively.
In our traffic-management system, we assume that each intersection controller is equipped with a sensor suite to monitor traffic flow.
We additionally assume intersection controllers are connected to a remote system (the traffic-management subsystem) that can modify their timer parameters.

\begin{figure}[b]
    \centering
    \includegraphics[width=.8\columnwidth]{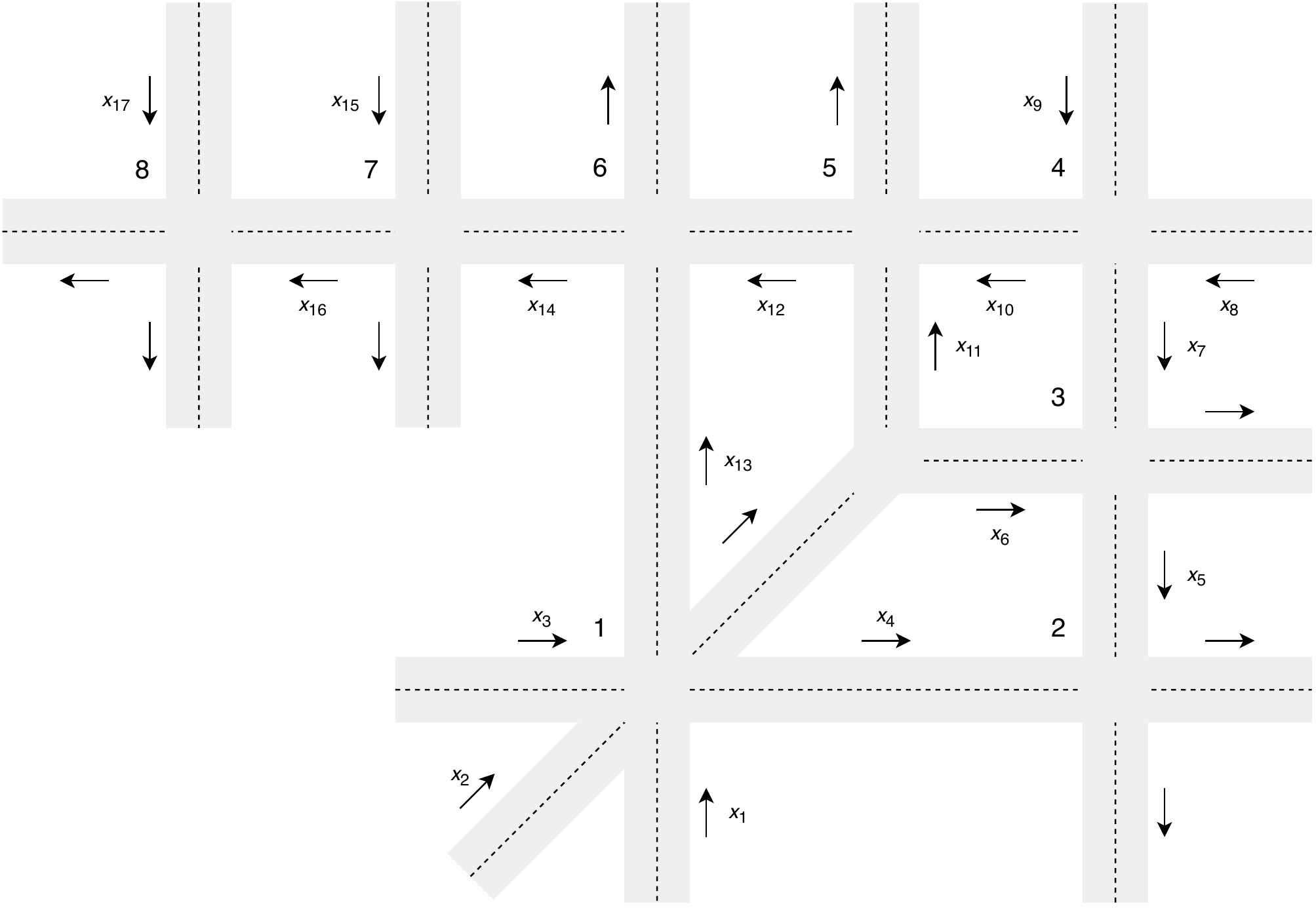}
    \caption{Traffic network.}
    \label{fig:traffic_network}
\end{figure}

For the computer-based simulations of the traffic-management system, we consider the following assumptions:
\begin{itemize}
\item
The traffic network is comprised of $17$ roads and $8$ intersections [10], which is shown in Fig. \ref{fig:traffic_network}.
\item
The traffic network states are the number of vehicles in the roads, and its inputs are the intersections' signal green-time [10].
\item
The simulation time is $4$ hours, with sampling rate of $10$ seconds.
\item
The cycle time for all the junctions is $120$ seconds.
\item
The vehicles' speed in the free flow rate is $40$ km/h.
\item
The traffic supply flow rate (demand) is a balanced Gaussian distribution function between $300$ and $1100$ vehicles. 
\item
The maximum and minimum values for the green time are $10$ and $600$ seconds, respectively (input constraints). 
\item
The capacity of each road, state constraint, is $1100$ vehicles.
\end{itemize}

\begin{figure}[t]
	\centering
    \includegraphics[width=\columnwidth]{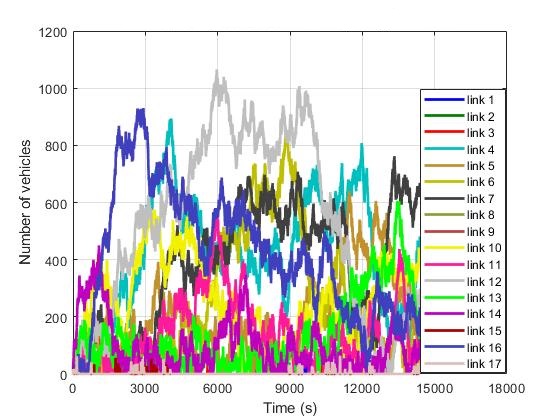}
	\caption{17 road state trajectories using the distributed model-based controller}
	\label{fig:states}
\end{figure}

Fig. \ref{fig:states} shows the $17$ road state trajectories of simulated traffic management system. Based on the state graphs, the traffic management control strategy shows satisfactory performance in optimizing the number of vehicles in each road. The controllers distribute the traffic congestion in the roads evenly, and perfectly satisfy the system state and control input constraints. 

The communication effort and therefore the computational overhead is significantly low using the proposed distributed scheme. 
Moreover, the controllers are designed fault-tolerant; e.g. a failure in one subsystem or sub-controller can be handled in the distributed fault-free control scheme [12]. 

\subsection{Data Analytics System and Database}
The data analytics system is implemented as a Python server operating in a virtual environment.
In its current form, the server listens for incoming data from the connected vehicles and intersection controllers.
The data for each vehicle is stored in a PostgreSQL database operating on another server in a separate virtual environment.

Using this data, the analytics system can learn trends or patterns that are in turn used to update the existing models for the testbed's systems.
Additionally, the analytics system can create data abstractions or inferences to pass to the dashboard.

\subsection{Dashboard}
The dashboard is the main system interface for human operators. 
It is implemented as a Python-based web server operating in a virtual environment.
When users access the dashboard website, they are presented with graphs showing system changes over time.
Additionally, the current state of all connected systems is displayed.

The dashboard retrieves data from the database using the PyMySQL application programming interface (API) and displays it using the Dash framework by Plotly.
Changes made to system parameters by users generate events that are then propagated through the network to the appropriate system or controller.

\subsection{Networking}
The backbone networking infrastructure is responsible for communication among all parts of the testbed.
All of the servers communicate over a 10 Gb Ethernet connection to form the core network.
Clients, such as the vehicles and intersection controllers, communicate over either a 2.4 GHz or 5 GHz Wi-Fi connection.

\section{Conclusion and Future Work}
\label{sec:conclusion}
Intelligent transportation systems are quickly increasing in both scale and complexity, and simulation software is unable to provide adequate fidelity.
We therefore proposed a physical testbed for traffic management and vehicle control as part of a larger smart city testbed.
Our system can be used to research a wide variety of intelligent transportation and vehicle control systems.
Additionally, it can be used as an educational tool for students interested in learning more about these systems.

Because implementing our testbed is still in its preliminary stages, our future plans involve a variety of tasks:
\begin{itemize}
    \item Increasing the size of both the intersection controller network and the vehicle network.
    \item Finalizing the networking infrastructure to support the traffic-management system
    \item Integrating an indoor positioning system, such as the OptiTrack motion capture system, to capture absolute positions of vehicles.
    \item Increasing the capabilities of the dashboard
\end{itemize}
Once complete, this testbed will be merged with a smart building testbed [19] to provide a complete platform for smart-city research.

\end{document}